\pdfoutput=1

\documentclass[]{INCLUDES/llncs}
\usepackage{graphicx}
\usepackage{subfig}
\usepackage{url}
\usepackage{verbatim} 
\usepackage{listings}
\lstnewenvironment{codex}
    {\lstset{}%
      \csname lst@SetFirstLabel\endcsname}
    {\csname lst@SaveFirstLabel\endcsname}
\lstnewenvironment{spec}
    {\lstset{}%
      \csname lst@SetFirstLabel\endcsname}
    {\csname lst@SaveFirstLabel\endcsname}
\lstnewenvironment{code}
    {\lstset{}%
      \csname lst@SetFirstLabel\endcsname}
    {\csname lst@SaveFirstLabel\endcsname}
\newtheorem{prop}{Proposition}

\newtheorem{df}{Definition}

\newcommand{\BI}[0]{\begin{itemize}}
\newcommand{\EI}[0]{\end{itemize}}

\newcommand{\BE}[0]{\begin{enumerate}}
\newcommand{\EE}[0]{\end{enumerate}}

\newcommand{\BX}[0]{\begin{codex}}
\newcommand{\EX}[0]{\end{codex}}
\newcommand{\BV}[0]{\begin{verbatim}}
\newcommand{\EV}[0]{\end{verbatim}}
    
\def \bscale1 {0.25}
\def \bscale {0.25}

\title{
  A Unified Formal Description of Arithmetic and Set Theoretical Data Types
}
\author{Paul Tarau}
\institute{
   Department of Computer Science and Engineering\\
   University of North Texas\\
   {\em E-mail: tarau@cs.unt.edu}
}

\begin{document}
\maketitle
\date{}

\begin{abstract}

We provide a ``shared
axiomatization'' of natural numbers and
hereditarily finite sets built around a polymorphic
abstraction of bijective base-2 arithmetics.

The ``axiomatization'' is described as a progressive refinement of Haskell type
classes with examples of instances converging to an efficient implementation in
terms of arbitrary length integers and bit operations.
As an instance, we derive algorithms to perform arithmetic
operations efficiently directly with hereditarily finite sets.

The self-contained source code of the paper is available at
\url{http://logic.cse.unt.edu/tarau/research/2010/unified.hs} .

{\bf Keywords:} {\em
formal description of arithmetic and set theoretical data types,
Peano arithmetic and hereditarily finite sets,
bijective base-2 arithmetic,
software refinement with Haskell type classes,
computational mathematics
%%
\begin{comment}
We define type classes that unify arithmetic and finite set theory.
As a result, arithmetic operations can be performed efficiently 
with trees representing hereditarily finite sets and a total order on
finite sets shading new light on the bi-interpretability between
Peano arithmetic and 
Zermelo-Fraenkel set theory with the negation of the
axiom of infinity and $\epsilon$-induction is derived.
\end{comment}
% finite set theory and arithmetic
% modeling axiom systems in functional programming
% symbolic arithmetic with tree representations
% software refinement with Haskell type classes
% executable theoretical computer science
}
\end{abstract}

\section{Introduction}

Natural numbers and finite sets have been used as sometimes competing
foundations for mathematics, logic and consequently computer science. The
de facto standard axiomatization for natural numbers is provided
Peano arithmetic. 
Finite set theory is axiomatized with the usual
Zermelo-Fraenkel system (abbreviated $ZF$) 
in which the Axiom of Infinity is
replaced by its negation.
When the axiom of $\epsilon$-induction, (saying that
if properties proven on elements also hold on sets containing them, then
they hold for all finite sets) is added, the resulting finite set theory 
(abbreviated $ZF^*$) is {\em bi-interpretable} with Peano arithmetic
i.e. they emulate each other accurately through a bijective mapping 
that commutes with standard operations on the two sides (\cite{kaye07}).

{\em
This foundational convergence suggests a ``shared
axiomatization'' of Peano arithmetic,
hereditarily finite sets and 
more conventional natural number representations
to be used as a unified framework for
formally deriving various computational entities.}

While axiomatizations of various formal systems are traditionally expressed in
classic or intuitionistic predicate logic, equivalent formalisms, in particular
the $\lambda$-calculus and the type theory used in modern functional
languages like Haskell, can provide specifications in a sometime more
readable, more concise, and more importantly, 
in a genuinely {\em executable} form.

Our incremental specification loop consists of successive refinements
through a chain of Haskell {\em type classes} (seen as axiom systems)
connected by inheritance.
 
{\em Instances} of the type classes (seen as {\em interpretations} of 
axiom systems) provide examples
that implement various data types in this framework.

The resulting hierarchy of type classes describes incrementally
{\em common computational capabilities} shared by bit-stacks, 
Peano natural numbers and
hereditarily finite sets   (sections
\ref{sharing}-\ref{arithops}).

\section{Computing in bijective base-2} \label{bstack}
  
Bitstrings provide a common and efficient 
computational representation for both
sets and natural numbers. 
This  recommends their operations as the 
right abstraction
for deriving, 
in the form of a Haskell type class, a 
``shared axiomatization''
for Peano arithmetic and Finite Set Theory.

While the existence of such a common axiomatization can be seen as a
consequence of the bi-interpretability results proven in \cite{kaye07}, our
distinct executable specification as a Haskell type class provides 
unique insights into the shared
inductive constructions and ensures that computational complexity of operations
is kept under control for a variety of instances.

We start by expressing bitstring operations as a Haskell data type:
\begin{comment}
\begin{code}
module Unified where
import Data.List
import Data.Bits
\end{code}
\end{comment}

\begin{code}
data BitStack = Empty|Bit0 BitStack|Bit1 BitStack 
  deriving (Eq, Show, Read)
\end{code}
We define the following operations on BitStacks
\begin{code}
empty = Empty

pushBit0 xs = Bit0 xs  
pushBit1 xs = Bit1 xs

popBit (Bit0 xs)=xs
popBit (Bit1 xs)=xs
\end{code}
and the predicates
\begin{code}
empty_ x=Empty==x
bit0_ (Bit0 _)=True
bit0_ _ =False

bit1_ (Bit1 _)=True
bit1_ _=False
\end{code}

We remind a few basic (but possibly not widely known) concepts related
to the computation mechanism we will use on 
bitstrings\footnote{We assume that bitstrings are mapped to numbers 
starting with the lowest exponent of 2 
and ending with  the highest.}.
\begin{df}
Bijective base-2 representation associates to $n \in \mathcal{N}$ a unique 
string in the regular language
$\{0,1\}^*$ by removing the 1 indicating the highest exponent of 2 
from the standard (complement of 2) bitstring representation of $n+1$.
\end{df}
Using a list notation for bitstrings this gives: 
$0=[], 1=[0], 2=[1], 3=[0,0], 4=[1,0], 5=[0,1], 6=[1,1]$ 
etc\footnote{See 
\url{http://en.wikipedia.org/wiki/Bijective_numeration} for the
historical origins of the concept and the 
more general {\em bijective base-k} case.}.

As a simple exercise in bijective base-2, 
arithmetic
one can now implement the successor function - and therefore provide a
model of Peano's axioms, as follows:
\begin{code}
zero = empty
one = Bit0 empty  
  
peanoSucc xs | empty_ xs = one
peanoSucc xs | bit0_ xs = pushBit1 (popBit xs)
peanoSucc xs | bit1_ xs = pushBit0 (peanoSucc (popBit xs)) 
\end{code}
For instance, 3 applications of {\tt peanoSucc}
generate $3=[0,0]$ as follows:
\begin{codex}
*Unified> (peanoSucc . peanoSucc . peanoSucc) zero
Bit0 (Bit0 Empty)
\end{codex}
One can verify by structural induction that:
\begin{prop}
Peano's axioms hold with the definition of the successor function provided by 
{\tt peanoSucc}.
\end{prop}

Using the BitStack representation (by contrast
with naive ``base-1" successor based definitions),
one can implement arithmetic operations like
sum and product with low polynomial complexity in terms of the bitsize of
their operands. 
We will defer defining these operations until the next sections, 
where we will provide such implementations in a more general setting.
 
Note that as a mild lookahead step towards abstracting away operations on our
bitstacks, we have replaced reference to data constructors
by the corresponding predicates and 
functions i.e. {\tt bit0\_} {\tt bit1\_} etc. 

\begin{comment}
We will spare the kind reader from a similar exercise showing basic set
operations on our bitstacks seen as characteristic functions of sets, and just
conclude this section by saying, that in a nutshell, our bitstacks 
promise to have the capabilities
needed to emulate both Peano arithmetic and ZF-finite sets in a single 
framework.
\end{comment}

\section{Sharing axiomatizations with {\em type classes}} \label{sharing}

Haskell's {\em type classes} \cite{Jones97typeclasses} are 
a good approximation of axiom
systems as they allow one to describe properties and operations 
generically i.e. in
terms of their action on objects of a parametric type. Haskell's {\em instances}
approximate {\em interpretations} \cite{kaye07} of such axiomatizations by
providing implementations of primitive operations and by refining and possibly
overriding derived operations with more efficient equivalents.

We will start by defining a type class that abstracts away the operations on the
{\tt BitStack} datatype and provides an axiomatization of natural numbers first,
and hereditarily finite sets later. 

\subsection{The 5 primitive operations}

The class {\tt Polymath} assumes only a theory of structural equality (as
implemented by the class {\tt Eq} in Haskell) and the {\tt Read/Show} 
superclasses needed for input/output. 

An instance of this class is required to implement the following 5
\begin{comment}
\footnote{As Poincar\'{e} has noticed it long time ago, talking
about {\tt 5} at this point is inherently circular: we are just about to define
what numbers are - and {\tt 5} is one of them. The standard excuse is that we
are not alone - the comment applies generically to the foundations shared by
logic, mathematics and computer science where practitioners are eager to
ignore it under the claim that it happens by confusing levels of abstraction.}
\end{comment}
primitive operations:
\begin{code}
class (Eq n,Read n,Show n)=>Polymath n where  
  e :: n 
  o_ :: n->Bool
  o :: n->n 
  i :: n->n
  r :: n->n
\end{code}
We have chosen single letter names {\tt e,o\_,o,i,r}
for the abstract operations corresponding respectively 
to {\tt empty, bit0\_, pushBit0, pushBit1, popBit} to facilitate a 
concise ``algebraic'' view needed to grasp some complex definitions 
that use compositions of these operations\footnote{ 
As an ongoing analogy,
the reader can interpret {\tt o} as pushing a {\tt 0} 
to a bitstack, {\tt i} as pushing a {\tt 1} and {\tt r} as a pop operation, 
with {\tt e} representing
an empty bitstack.}. 

The {\tt Polymath} type class also provides to its instances 
generic implementations of
the following derived operations:
\begin{code}
  e_ :: n->Bool
  e_ x = x==e

  i_ :: n->Bool
  i_ x = not (o_ x || e_ x)
\end{code}
Note that we use the convention that for each constructor 
the recognizer's name is obtained by appending ``{\tt \_}"\footnote{
As part of the bitstack analogy, the predicates
{\tt o\_} and {\tt i\_} can be seen as recognizing respectively a {\tt 0}
and a {\tt 1} (in bijective base-2) at the top of the  bitstack.}.

While not strictly needed at this point, it is convenient also to include in
the {\tt Polymath} type class some additional derived operations.
As we will see later, some instances will chose to override them.
We first define
an object and a recognizer for {\tt 1}, 
the constant function {\tt u} and the
predicate {\tt u\_}.
\begin{code}
  u :: n
  u = o e
  
  u_ :: n->Bool
  u_ x = o_ x && e_ (r x)
\end{code}
Next we implement the successor {\tt s} and predecessor {\tt p} functions:
\begin{code}  
  s :: n->n
  s x | e_ x = u
  s x | o_ x = i (r x)
  s x | i_ x = o (s (r x)) 
  
  p :: n->n
  p x | u_ x = e
  p x | o_ x = i (p (r x)) 
  p x | i_ x = o (r x)
\end{code}
It is convenient at this point, as we target a diversity of interpretations
materialized as Haskell instances, to provide a polymorphic converter between
two different instances of the type class {\tt Polymath}.
The function {\tt view} allows converting between two
different Polymath instances, generically.
\begin{code}
view :: (Polymath a,Polymath b)=>a->b
view x | e_ x = e
view x | o_ x = o (view (r x))
view x | i_ x = i (view (r x))
\end{code}

\subsection{Peano arithmetic}
It is important to observe at this point that Peano arithmetic is an
instance of the class {\tt Polymath} i.e. that the class can be used to
derive an ``axiomatization'' for Peano arithmetic
through a straightforward
mapping of Haskell's function definitions 
to Peano's axioms.
\begin{code}
data Peano = Zero|Succ Peano deriving (Eq,Show,Read)

instance Polymath Peano where
  e = Zero
  
  o_ Zero = False
  o_ (Succ x) = not (o_ x) 
  
  o x = Succ (o' x) where
    o' Zero = Zero
    o' (Succ x) = Succ (Succ (o' x))
    
  i x = Succ (o x)
  
  r (Succ Zero) = Zero
  r (Succ (Succ Zero)) = Zero
  r (Succ (Succ x)) = Succ (r x) 
\end{code}
Finally, we can add {\tt BitStack} - which, after all, has inspired the
operations of our type class, as an instance of {\tt Polymath}
\begin{code}
instance Polymath BitStack where
  e=empty
  o=pushBit0
  o_=bit0_ 
  i=pushBit1
  r=popBit
\end{code}
and observe that the Peano and Bitstack interpretations behave consistently:
\begin{codex}
*Unified> i (o (o Empty))
Bit1 (Bit0 (Bit0 Empty))
*Unified> i (o (o Zero))
Succ (Succ (Succ (Succ (Succ (Succ (Succ (Succ Zero)))))))
*Unified> i (o (o Empty))
Bit1 (Bit0 (Bit0 Empty))
*Unified> s it
Bit0 (Bit1 (Bit0 Empty))
*Unified> view it :: Peano
Succ (Succ (Succ (Succ (Succ (Succ (Succ (Succ (Succ Zero))))))))
*Unified> p it
Succ (Succ (Succ (Succ (Succ (Succ (Succ (Succ Zero)))))))
Bit1 (Bit0 (Bit0 Empty))
\end{codex}
Note also the convenience of using {\tt :: view} to instantly
morph between instances and the use of Haskell's {\tt it} standing for the
previously returned result.
So far we have seen that our instances implement
syntactic variations of natural numbers equivalent to
Peano's axioms.
We will now provide an instance showing that our ``axiomatization'' covers
the theory of hereditarily finite sets (assuming, of course, that
extensionality, comprehension, regularity, $\epsilon$-induction etc. are
implicitly provided by type classes like {\tt Eq} and 
implementation of recursion in the underlying
programming language).

\section{Computing with {\em hereditarily finite sets}} \label{chfs}

Hereditarily finite sets are built inductively from the empty set (denoted {\tt
S []}) by adding finite unions of existing sets at each stage. We first define a
rooted tree datatype {\tt S}:
\begin{code}
data S=S [S] deriving (Eq,Read,Show)
\end{code}
To accurately represent sets, the type {\tt S} would require a type system
enforcing constraints on type parameters, saying that all elements
covered by the definition are distinct and no repetitions occur
in any list of type {\tt [S]}. We will assume this and similar properties
of our datatypes, when needed, from now on, and consider trees built with
the constructor {\tt S} as representing hereditarily finite sets.

We will now show that hereditarily finite sets can do arithmetic
as instances of the class {\tt Polymath} by implementing a successor (and
predecessor) function. We start with the easier operations:
\begin{code}
instance Polymath S where
  e = S []
  
  o_ (S (S []:_)) = True
  o_ _ = False
  
  o (S xs) = s (S (map s xs))
  
  i = s . o
\end{code}
Note that the {\tt o} operation, that can be seen as pushing a {\tt 0} bit to
a bitstack is implemented by applying {\tt
s} to each branch of the tree. We will now implement {\tt r, s} and {\tt p}.
\begin{code}  
  r (S xs) = S (map p (f ys)) where 
    S ys = p (S xs)
    f (x:xs) | e_ x = xs
    f xs = xs
   
  s (S xs) = S (hLift (S []) xs) where
    hLift k [] = [k]
    hLift k (x:xs) | k==x = hLift (s x) xs
    hLift k xs = k:xs

  p (S xs) = S (hUnLift xs) where
    hUnLift ((S []):xs) = xs
    hUnLift (k:xs) = hUnLift (k':k':xs) where k'= p k 
\end{code}
First note that {\em successor} and {\em predecessor} operations {\tt s,p} are overridden
and that the {\tt r} operation is expressed in terms of {\tt p}, as {\tt o} and
{\tt i} were expressed in terms of {\tt s}.
Next, note that the {\tt map} combinators 
and the auxiliary functions {\tt hLift} and {\tt hUnlift} 
are used to delegate work between successive levels of the 
tree defining a hereditarily finite set.

To summarize, let us observe that the successor and predecessor 
operations {\tt s,p} at a given level are implemented
through iteration of the same at a lower level and that the
``left shift'' operation implemented by {\tt o,i} results 
in initiating {\tt s} operations at a lower level. 
Thus the total number of
operations is within a constant factor of the size of the trees.

Let us verify that these operations mimic indeed their more common
counterparts on type {\tt Peano}.
\begin{codex}
*Unified> o (i (S []))
S [S [],S [S [S []]]]
*Unified> s it
S [S [S []],S [S [S []]]]
*Unified> view it :: Peano
Succ (Succ (Succ (Succ (Succ (Succ Zero)))))
*Unified> p it
Succ (Succ (Succ (Succ (Succ Zero))))
*Unified> view it :: S
S [S [],S [S [S []]]]
\end{codex}

It can be proven by structural induction that:
\begin{prop}
Hereditarily finite sets as represented by the data type {\tt S} 
implement the same successor and predecessor operation as 
the instance {\tt Peano}.
\end{prop}
Note that this implementation of the class {\tt Polymath}
implicitly uses the {\em Ackermann interpretation} of Peano arithmetic in terms of
the theory of hereditarily finite sets, i.e. the natural number associated
to a hereditarily finite set is given by the function
\vskip 0.30cm
$f(x)$ = {\tt if} $x=\emptyset$ {\tt then} $0$ {\tt else} $\sum_{a \in
x}2^{f(a)}$ 
\vskip 0.30cm

Let us summarize what's unusual with instance {\tt S} of the class {\tt
Polymath}: it shows that successor and predecessor operations can be 
performed with {\em hereditarily finite
sets playing the role of natural numbers}. As natural numbers and finite
ordinals are in a one-to-one mapping, this instance shows that hereditarily
finite sets can be seen as {\em finite ordinals} directly, without using the simple
but computationally explosive von Neumann construction (which defines ordinal
$n$ as the set $\{0,1,\ldots,n-1\}$). We will elaborate more on this
after defining a total order on our Polymath type.

\section{Arithmetic operations} \label{arithops}
Our next refinement adds key arithmetic operations in the form
of a type class extending {\tt Polymath}. 
We start with addition ({\tt polyAdd}) and subtraction ({\tt polySubtract}):
\begin{code} 
class (Polymath n) => PolyOrd n where
  polyAdd :: n->n->n 
  polyAdd x y | e_ x = y
  polyAdd x y | e_ y = x
  polyAdd x y | o_ x && o_ y =    i (polyAdd (r x) (r y))
  polyAdd x y | o_ x && i_ y = o (s (polyAdd (r x) (r y)))
  polyAdd x y | i_ x && o_ y = o (s (polyAdd (r x) (r y)))
  polyAdd x y | i_ x && i_ y = i (s (polyAdd (r x) (r y)))
  
  polySubtract :: n->n->n
  polySubtract x y | e_ x && e_ y = e
  polySubtract x y | not(e_ x) && e_ y = x
  polySubtract x y | not (e_ x) && x==y = e
  polySubtract z x | i_ z && o_ x = o (polySubtract (r z) (r x))  
  polySubtract z x | o_ z && o_ x = i (polySubtract (r z) (s (r x)))  
  polySubtract z x | o_ z && i_ x = o (polySubtract (r z) (s (r x)))
  polySubtract z x | i_ z && i_ x = i (polySubtract (r z) (s (r x)))  
\end{code}
Efficient comparison uses the fact that with our representation
only sequences of distinct lengths can be different. We start by
comparing lengths:
\begin{code}
  lcmp :: n->n->Ordering
  
  lcmp x y | e_ x && e_ y = EQ 
  lcmp x y | e_ x && not(e_ y) = LT
  lcmp x y | not(e_ x) && e_ y = GT
  lcmp x y = lcmp (r x) (r y)
\end{code}
Comparison can now proceed by case analysis, the interesting
case being when lengths are equal (function {\tt samelen\_cmp}):
\begin{code} 
  cmp :: n->n->Ordering
  cmp x y = ecmp (lcmp x y) x y where
     ecmp EQ x y = samelen_cmp x y
     ecmp b _ _ = b
     
  samelen_cmp :: n->n->Ordering

  samelen_cmp x y | e_ x && e_ y = EQ
  samelen_cmp x y | e_ x && not(e_ y) = LT
  samelen_cmp x y | not(e_ x) && e_ y = GT
  samelen_cmp x y | o_ x && o_ y = samelen_cmp (r x) (r y)
  samelen_cmp x y | i_ x && i_ y = samelen_cmp (r x) (r y)
  samelen_cmp x y | o_ x && i_ y = 
    downeq (samelen_cmp (r x) (r y)) where
      downeq EQ = LT
      downeq b = b
  samelen_cmp x y | i_ x && o_ y = 
    upeq (samelen_cmp (r x) (r y)) where
      upeq EQ = GT
      upeq b = b
\end{code}
Finally, boolean comparison operators are defined as follows:
\begin{code}
  lt,gt,eq :: n->n->Bool
  
  lt x y = LT==cmp x y
  
  gt x y = GT==cmp x y
  
  eq x y = EQ==cmp x y
\end{code}
\begin{comment}
We are now ready for a sorting operation, derived from Haskell's parametric {\tt
sortBy}. We define our sorting function {\tt polySort} as follows:
\begin{code}
  polySort :: [n]->[n]
  polySort ns = sortBy polyCompare ns
  
  polyCompare :: n->n->Ordering
  polyCompare x y | x==y = EQ
  polyCompare x y | lt x y = LT
  polyCompare _ _ = GT
\end{code}
\end{comment}
After adding the instances
\begin{code}
instance PolyOrd Peano
instance PolyOrd BitStack
instance PolyOrd S

\end{code}
one can see that all operations extend naturally:
\begin{codex}
*Unified> polyAdd (Succ Zero) (Succ Zero)
Succ (Succ Zero)
*Unified> (s.s.s.s) Empty
Bit1 (Bit0 Empty)
*Unified> take 1000 (iterate s (S []))
[S [],S [S []],....,S [S [],S [S [],S [S []]]]]]
*Unified> and (zipWith lt it (map s it))
True
\end{codex}
The last example confirms, for 1000 instances,
that we have a {\em well-ordering} of hereditarily
finite sets without recurse to the von Neumann ordinal construction (used in
\cite{kaye07} to complete the bi-interpretation from hereditarily finite sets
to natural numbers). This replicates a recent result described in
\cite{petti09} where a lexicographic ordering is used to simplify the
proof of bi-interpretability of \cite{kaye07}.

We will proceed now with introducing more powerful operations. Needless to say,
they will apply automatically to all instances of the type class {\tt Polymath}.

\section{Adding other arithmetic operations} \label{morearith}
We first define multiplication.
\begin{code}
class (PolyOrd n) => PolyCalc n where
  polyMultiply :: n->n->n
  polyMultiply x _ | e_ x = e
  polyMultiply _ y | e_ y = e
  polyMultiply x y = s (multiplyHelper (p x) (p y)) where
    multiplyHelper x y | e_ x = y
    multiplyHelper x y | o_ x = o (multiplyHelper (r x) y)
    multiplyHelper x y | i_ x = s (polyAdd y  (o (multiplyHelper (r x) y)))
  
  double :: n->n
  double = p . o
  
  half :: n->n
  half = r . s
\end{code}
Exponentiation by squaring follows - easier for powers of two ({\tt exp2}), then
the general case ({\tt pow}):
\begin{code}
  exp2 :: n->n -- power of 2
  exp2 x | e_ x = u
  exp2 x = double (exp2 (p x)) 
  
  pow :: n->n->n -- power y of x
  pow _ y | e_ y = u
  pow x y | o_ y = polyMultiply x (pow (polyMultiply x x) (r y))
  pow x y | i_ y = polyMultiply 
    (polyMultiply x x) 
    (pow (polyMultiply x x) (r y)) 
\end{code}    
After defining instances
\begin{code}  
instance PolyCalc Peano
instance PolyCalc BitStack
instance PolyCalc S
\end{code}
operations can be tested under various representations
\begin{codex}
*Unified> polyMultiply (s (s (S []))) (s (s (s (S []))))
S [S [S []],S [S [S []]]]
*Unified> view it :: Peano
Succ (Succ (Succ (Succ (Succ (Succ Zero)))))
*Unified> pow (s (s (S []))) (s (s (s (S []))))
S [S [S [],S [S []]]]
*Unified> view it :: Peano
Succ (Succ (Succ (Succ (Succ (Succ (Succ (Succ Zero)))))))
\end{codex}

\section{Deriving set operations} \label{setops}
We will now provide
a set view of our polymorphic data type.
Following \cite{calc09fiso}, where Ackermann's mapping
between hereditarily finite sets and natural numbers
has been derived as a fold/unfold operation using a
bijection between natural numbers and finite sets of natural
numbers, we can write:
\begin{code}
class (PolyCalc n) => PolySet n where
  as_set_nat :: n->[n]
  as_set_nat n = nat2exps n e where
    nat2exps n _ | e_ n = []
    nat2exps n x = if (i_ n) then xs else (x:xs) where
      xs=nat2exps (half n) (s x)

  as_nat_set :: [n]->n
  as_nat_set ns = foldr polyAdd e (map exp2 ns)
\end{code}
Given that natural numbers and 
hereditarily finite sets, when seen as instances of our generic
axiomatization, are connected through Ackermann's bijections,
one can shift from one side to the other at will:
\begin{codex}
*Unified> as_set_nat (s (s (s Zero)))
[Zero,Succ Zero]
*Unified> as_nat_set it
Succ (Succ (Succ Zero))
*Unified> as_set_nat (s (s (s (S []))))
[S [],S [S []]]
*Unified> as_nat_set it
S [S [],S [S []]]
\end{codex}
Note also that, as the operations on type {\tt S} show, 
the set associated to the number 3 is exactly 
the same as the first level of
its expansion as a hereditarily finite set.

After defining combinators for operations of arity 1 and 2:
\begin{code}
  setOp1 :: ([n]->[n])->(n->n)
  setOp1 f = as_nat_set . f . as_set_nat 
  setOp2 :: ([n]->[n]->[n])->(n->n->n)
  setOp2 op x y = as_nat_set (op (as_set_nat x) (as_set_nat y))
\end{code}
we can ``borrow'' (with confidence!) the usual set operations (provided in
the Haskell package Data.List):
\begin{code}
  setIntersection :: n->n->n
  setIntersection = setOp2 intersect
                   
  setUnion :: n->n->n
  setUnion = setOp2 union
  
  setDifference :: n->n->n
  setDifference = setOp2 (\\)
  
  setIncl :: n->n->Bool
  setIncl x y = x==setIntersection x y
\end{code}
In a similar way, we define a powerset operation
conveniently using actual lists, before reflecting
it into an operation on natural numbers.
\begin{code}
  powset :: n->n
  powset x = as_nat_set 
    (map as_nat_set (subsets (as_set_nat x))) where
      subsets [] = [[]]
      subsets (x:xs) = [zs|ys<-subsets xs,zs<-[ys,(x:ys)]]   
\end{code}
Next, the $\epsilon$-relation defining set membership is
given as the function {\tt inSet}, together with the {\tt augmentSet} function
used in various set theoretic constructs as a new set generator.
\begin{code}
  inSet :: n->n->Bool
  inSet x y = setIncl (as_nat_set [x]) y 
  
  augmentSet :: n->n
  augmentSet x = setUnion x (as_nat_set [x])
\end{code} 
The $n$-th {\em von Neumann ordinal}
is the set $\{0,1,\ldots,n-1\}$
and it is used to emulate natural numbers
in finite set theory.
It is implemented by the 
function {\tt nthOrdinal}:
\begin{code}
  nthOrdinal :: n->n
  nthOrdinal x | e_ x = e
  nthOrdinal n = augmentSet (nthOrdinal (p n)) 
\end{code}
Note that as hereditarily finite sets and natural numbers 
are instances of the class {\tt PolyOrd}, an order
preserving bijection can be defined between the two,
which makes it unnecessary to resort to von Neumann ordinals
to show bi-interpretability \cite{kaye07,petti09}.

After defining the appropriate instances
\begin{code}
instance PolySet Peano
instance PolySet BitStack
instance PolySet S
\end{code}
we observe that set operations act naturally under the
hereditarily finite set interpretation:
\begin{codex}
*Unified> (s.s.s.s.s.s)  (S [])
S [S [S []],S [S [S []]]]
*Unified> inSet (S [S []]) it
True

*Unified> powset (S [])
S [S []]
*Unified> powset it
S [S [],S [S []]]

*Unified> augmentSet (S [])
S [S []]
*Unified> augmentSet it
S [S [],S [S []]]
\end{codex}

\section{Deriving an instance with fast bitstring operations} \label{fastops}

We will now benefit from our shared axiomatization
by designing an instance that takes advantage of bit operations, to implement,
through a few overrides, fast versions of our arithmetic and set functions.
For syntactic convenience, we will map this instance directly to Haskell's
arbitrary length Integer type, to benefit in GHC from the performance of
the underlying C-based GMP package. First some arithmetic operations
(making use of Haskell's {\tt Data.Bits} library):
\begin{code}
instance Polymath Integer where
  e = 0
  o_ x = testBit x 0
 
  o x = succ (shiftL x 1) 
  i  = succ . o
  r x | x>0 = shiftR (pred x) 1

  s = succ
  p n | n>0 = pred n
  u = 1
  u_ = (== 1)
  
instance PolyOrd Integer where
  polySubtract x y = abs (x-y)
  lt = (<)
  polyCompare=compare

instance PolyCalc Integer where
  polyMultiply = (*)
  half x = shiftR x 1
  double x = shiftL x 1
\end{code}
Next, some set operations:
\begin{code}
instance PolySet Integer where
  setUnion = (.|.)
  setIntersection = (.&.)
  setDifference x y = x .&. (complement y)
  
  inSet x xs = testBit xs (fromIntegral x)
 
  powset 0 = 1
  powset x = xorL (powset (pred x)) where
    xorL n = n `xor` (shiftL n 1)
\end{code}
It is tempting to test for correctness, by 
computing with the ``implementation'' provided by the type 
{\tt Integer} and then reverting to the set view:
\begin{codex}
*Unified> as_nat_set [1,3,4]
26
*Unified> powset it
84215045
*Unified> map as_set_nat (as_set_nat it)
[[],[1],[3],[1,3],[4],[1,4],[3,4],[1,3,4]]
\end{codex}
It all adds up, but as we do not have a proof yet, 
we leave it as an {\em open problem} 
to show that {\em the {\tt xor} based
instance of {\tt powset} in {\tt Integer}
does indeed implement the powerset operation
as specified in section \ref{setops}}.

Finally, we can observe that the von Neumann ordinal construction
(used to introduce natural numbers in set theory) defines
a fast growing injective function from $\mathcal{N} \to  \mathcal{N}$:
\begin{codex}
*Unified> map nthOrdinal [0..4]
[0,1,3,11,2059]
*Unified> as_set_nat 2059
[0,1,3,11]
\end{codex}
In contrast, our ``shared axiomatization'' defines ordinals
through a trivial {\em bijection}: the identity function.

Note, as a more practical outcome, that one can 
now use arbitrary length integers
as an efficient representation of hereditarily finite sets.
Conversely, a computation like
\begin{codex} 
*Unified> s (S [S [S [S [S [S [S [S [S [S []]]]]]]]]])
S [S [],S [S [S [S [S [S [S [S [S []]]]]]]]]]
\end{codex}
computing easily the successor of a tower of exponents of 2, in terms of
hereditarily finite sets, would overflow any computer's memory when 
using a conventional integer representation.

\section{Related work} \label{related}

The techniques described in this paper
originate in the data transformation
framework described in
\cite{sac09fISO,calc09fiso,ppdp09pISO}.
The main new contribution is that while 
our previous work can be seen as
``an existence proof'' that, for instance, arithmetic
computations can be performed with symbolic objects
like hereditarily finite sets, here we
show it constructively. Moreover, we lift our
conceptual framework to 
a polymorphic axiomatization which turns out
to have as {\em interpretations} (instances in Haskell parlance)
natural numbers, bitstacks and hereditarily
finite sets.

Natural number encodings of hereditarily finite sets have 
triggered the interest of researchers in fields like 
Axiomatic Set Theory and Foundations of Logic
\cite{kaye07,DBLP:journals/mlq/Kirby07}.
%An emulation of Peano and conventional binary arithmetic operations
%in Prolog, is described in \cite{DBLP:conf/flops/KiselyovBFS08}.
A number of papers of J. Vuillemin develop similar techniques 
aiming to unify various data types, with focus on theories of boolean
functions and arithmetic \cite{DBLP:conf/birthday/Vuillemin03}.
Binary number-based axiomatizations of natural number arithmetic are likely to
be folklore, but having access to the 
the underlying theory of the calculus of constructions
\cite{coquand:calculus:ic:88} and the inductive proofs of
their equivalence with Peano arithmetic 
in the libraries of the {\tt Coq} 
\cite{Coq:manual} proof assistant has been
particularly enlightening to the author. On the other hand we 
have not found in the literature any
axiomatizations in terms of hereditarily finite sets,
as derived in this paper.
Future work is planned in proving with Coq
the equivalence of operations in Peano arithmetic with 
their counterparts in
the set theoretic interpretation of our type classes.

\section{Conclusion} \label{concl}
In the form of a literate Haskell program, we have built ``shared
axiomatizations'' of finite arithmetic and
hereditarily finite sets
using successive refinements of type classes.
%The derivation of successive extensions as Haskell type classes,
%has shown the benefits of polymorphism, combinators 
%together with a flexible ``object oriented'' coding style.

We have derived some unusual algorithms,
for instance, by expressing arithmetic
computations symbolically, in terms of
hereditarily finite sets. We have also provided
a well-ordering for hereditarily finite
sets that maps them to ordinals directly,
without using the von Neumann construction.

This has been made possible by
extending the techniques introduced in \cite{sac09fISO,calc09fiso,ppdp09pISO}
that allow observing the internal working of intricate
mathematical concepts through isomorphisms transporting
operations between fundamental data types.

\bibliographystyle{INCLUDES/splncs}

%\bibliography{INCLUDES/theory,tarau,INCLUDES/proglang,INCLUDES/biblio,INCLUDES/syn}

\begin{thebibliography}{}

\end{thebibliography}


\begin{thebibliography}{10}

\bibitem{kaye07}
Kaye, R., Wong, T.L.:
\newblock {On Interpretations of Arithmetic and Set Theory}.
\newblock Notre Dame J. Formal Logic Volume \textbf{48}(4) (2007)  497--510

\bibitem{Jones97typeclasses}
Jones, S.P., Jones, M., Meijer, E.:
\newblock Type classes: An exploration of the design space.
\newblock In: Haskell Workshop. (1997)

\bibitem{petti09}
Pettigrew, R.:
\newblock {On Interpretations of Bounded Arithmetic and Bounded Set Theory}
  Notre Dame J. Formal Logic Volume 50, Number 2 (2009), 141-151.

\bibitem{calc09fiso}
Tarau, P.:
\newblock {A Groupoid of Isomorphic Data Transformations}.
\newblock In Carette, J., Dixon, L., Coen, C.S., Watt, S.M., eds.: {Intelligent
  Computer Mathematics, 16th Symposium, Calculemus 2009, 8th International
  Conference MKM 2009 }, Grand Bend, Canada, Springer, LNAI 5625 (July 2009)
  170--185

\bibitem{sac09fISO}
Tarau, P.:
\newblock {Isomorphisms, Hylomorphisms and Hereditarily Finite Data Types in
  Haskell}.
\newblock In: {Proceedings of ACM SAC'09}, Honolulu, Hawaii, ACM (March 2009)
  1898--1903

\bibitem{ppdp09pISO}
Tarau, P.:
\newblock {An Embedded Declarative Data Transformation Language}.
\newblock In: {Proceedings of 11th International ACM SIGPLAN Symposium PPDP
  2009}, Coimbra, Portugal, ACM (September 2009)  171--182

\bibitem{DBLP:journals/mlq/Kirby07}
Kirby, L.:
\newblock {Addition and multiplication of sets}.
\newblock Math. Log. Q. \textbf{53}(1) (2007)  52--65

\bibitem{DBLP:conf/birthday/Vuillemin03}
Vuillemin, J.:
\newblock Digital algebra and circuits.
\newblock In Dershowitz, N., ed.: Verification: Theory and Practice. Volume
  2772 of Lecture Notes in Computer Science., Springer (2003)  733--746

\bibitem{coquand:calculus:ic:88}
Coquand, T., Huet, G.:
\newblock The calculus of constructions.
\newblock Information and Computation \textbf{76}(2/3) (1988)  95--120

\bibitem{Coq:manual}
\mbox{The Coq development team}:
\newblock The Coq proof assistant reference manual.
\newblock LogiCal Project. (2004) Version 8.0.

\end{thebibliography}

\end{document}